\documentclass[fleqn,10pt]{SelfArx} 

\usepackage{lipsum} 

\definecolor{color1}{RGB}{0,0,90} 
\definecolor{color2}{RGB}{0,20,20} 

\usepackage{hyperref} 

\hypersetup{hidelinks,colorlinks,breaklinks=true,urlcolor=color2,citecolor=color1,linkcolor=color1,bookmarksopen=false,pdftitle={Title},pdfauthor={Author}}

\JournalInfo{PLOS ONE, in press} 
\Archive{} 

\PaperTitle{On the sympatric evolution and evolutionary stability of coexistence by relative nonlinearity of competition} 

\Authors{Florian Hartig\textsuperscript{1,2}*, Tamara M\"unkem\"uller\textsuperscript{3,4,1}, Karin Johst\textsuperscript{1}, Ulf Dieckmann\textsuperscript{5}} 
\affiliation{\textsuperscript{1}\textit{Department of Ecological Modelling, Helmholtz Centre for Environmental Research - UFZ, Leipzig, Germany}} 
\affiliation{\textsuperscript{2}\textit{Biometry and Environmental System Analysis, University of Freiburg, Freiburg, Germany}} 
\affiliation{\textsuperscript{3}\textit{Laboratoire d'\'{E}cologie Alpine, Universit\'{e} Grenoble-Alpes, Grenoble, France}} 
\affiliation{\textsuperscript{4}\textit{Laboratoire d'\'{E}cologie Alpine, Centre National de la Recherche Scientifique, Grenoble, France}} 
\affiliation{\textsuperscript{5}\textit{Evolution and Ecology Program, International Institute for Applied System Analysis (IIASA), Laxenburg, Austria}} 

\affiliation{*\textbf{Corresponding author}: florian.hartig@biom.uni-freiburg.de} 

\newcommand{\keywordname}{Keywords} 

\Abstract{If two species exhibit different nonlinear responses to a single shared resource, and if each species modifies the resource dynamics such that this favors its competitor, they may stably coexist. This coexistence mechanism, known as relative nonlinearity of competition, is well understood theoretically, but less is known about its evolutionary properties and its prevalence in real communities. We address this challenge by using adaptive dynamics theory and individual-based simulations to compare community stabilization and evolutionary stability of species that coexist by relative nonlinearity. In our analysis, evolution operates on the species' density-compensation strategies, and we consider a trade-off between population growth rates at high and low resource availability. We confirm previous findings that, irrespective of the particular model of density dependence, there are many combinations of overcompensating and undercompensating density-compensation strategies that allow stable coexistence by relative nonlinearity. However, our analysis also shows that most of these strategy combinations are not evolutionarily stable and will be outcompeted by an intermediate density-compensation strategy. Only very specific trade-offs lead to evolutionarily stable coexistence by relative nonlinearity. As we find no reason why these particular trade-offs should be common in nature, we conclude that the sympatric evolution and evolutionary stability of relative nonlinearity, while possible in principle, seems rather unlikely. We speculate that this may, at least in part, explain why empirical demonstrations of this coexistence mechanism are rare, noting, however, that the difficulty to detect relative nonlinearity in the field is an equally likely explanation for the current lack of empirical observations, and that our results are limited to communities with non-overlapping generations and constant resource supply. Our study highlights the need for combining ecological and evolutionary perspectives for gaining a better understanding of community assembly and biogeographic patterns.
}

\begin{document}

\flushbottom 

\maketitle 

\section*{Introduction} 

Understanding the evolution and maintenance of ecological diversity is a fundamental objective of ecological research. While the basic mechanisms of evolution have largely remained unchallenged since Darwin's foundational work, assessing the relative importance of different mechanisms known or conjectured to drive patterns of diversity and speciation remains among the most controversial questions in the field \cite{Coyne-Speciation-2004, Dieckmann-Adaptivespeciation-2004, Gavrilets-Fitnesslandscapesand-2004, Wiens-Nicheconservatismas-2010}. 
 
Classically, the maintenance of diversity was thought to be determined by niches and the associated principle of competitive exclusion. Niche differentiation was accordingly seen as the dominant process explaining the evolution of species and functional diversity \cite{Hutchinson-HomagetoSanta-1959}. Yet, this claim has early been challenged by the fact that a large number of species seem to be supported by the same environmental niche (e.g. in "the paradox of the plankton"; see \cite{Hutchinson-paradoxofplankton-1961}). In response to this challenge, a growing list of more complex coexistence mechanisms has been proposed, including biotic interactions such as conspecific negative density dependence \cite{Janzen-HerbivoresandNumber-1970, Packer-Soilpathogensand-2000, Bever-Rootingtheoriesof-2010, Mangan-Negativeplant-soilfeedback-2010}; dispersal-mediated mechanisms \cite{TILMAN-CompetitionAndBiodiversity-1994, Yu-Competition-ColonizationTrade-offIs-2001, Banitz-Clumpedversusscattered-2008, Dislich-Whatenablescoexistence-2010}; dynamic and spatial extensions of the classical resource niche, such as the spatial and temporal storage effect \cite{Crawley-Timingofdisturbance-2004, Adler-Climatevariabilityhas-2006}; the interplay of assortative mating and environmental heterogeneity \cite{MGonigle-Sexualselectionenables-2012}; as well as combinations of the former \cite{Johst-Evolutionofcomplex-1999, Berkley-Turbulentdispersalpromotes-2010, Santos-Neutralcommunitiesmay-2011}. It has even been proposed that stabilizing effects are altogether negligible for the maintenance of highly-diverse communities \cite{Hubbell-UnifiedNeutralTheory-2001}. 

All these mechanisms are plausible, and it is therefore an open empirical and theoretical question to assess to what extent and at which scales they contribute to the observed spatial and temporal patterns of local species occurrences. To shed light on this question, many studies have concentrated on ecological processes at the community scale, either by analyzing empirical patterns of species, traits, and phylogenies in space and time \cite{Gravel-Speciescoexistencein-2011}, or by means of theoretical models that explore the consequences of potential coexistence mechanisms. However, it has proven surprisingly difficult to arrive at an agreement even about fundamental issues with this approach, such as the extent to which non-neutral processes are responsible for the local structure of tropical plant communities (e.g. \cite{Chave-Neutraltheoryand-2004}). 

Evolutionary analyses might allow us to look at these questions from a new angle. Speciation and the functional divergence of species may occur due to random processes alone, but selection on ecological traits and functions in most cases seems to be a dominant driver \cite{Schluter-EvidenceEcologicalSpeciation-2009}. This suggests that looking at the plausibility of coexistence mechanisms from an evolutionary perspective might complement existing attempts to infer their importance from empirical data \cite{Hutchinson-EcologicalTheaterand-1969}. For example, Purves and Turnbull argue that it is highly unlikely that evolution would give rise to a large number of functional differences that are nevertheless perfectly fitness-equalizing \cite{Purves-Differentbutequal-2010}, a mechanism that has been suggested as an explanation for the neutral appearance of tropical plant communities (\cite{Hubbell-UnifiedNeutralTheory-2001}, see also the discussion in \cite{Kneitel-Trade-offsincommunity-2004}). Other recent studies have examined the conditions under which the storage effect is likely to evolve \cite{Snyder-CoexistenceandCoevolution-2011, Abrams-Evolutionstorageeffect-2013}. In general, however, there are still very few studies that connect evolutionary analyses with community-ecological questions, such as the relative importance of different assembly and coexistence mechanisms.

In this study, we apply an evolutionary rationale to relative nonlinearity of competition (RNC), a well-known dynamic coexistence mechanism \cite{Armstrong-CompetitiveExclusion-1980, Chesson-MultispeciesCompetitionin-1994}. RNC arises when species show different nonlinear responses to one or several common limiting factors, and each species affects the availability or fluctuations of those factors in a way that it decreases its own fitness when it becomes abundant \cite{Chesson-Mechanismsofmaintenance-2000}. This endogenous control of the resource dynamics is the main difference to the storage effect, the other commonly discussed coexistence mechanism in fluctuating environments. The theoretical properties of RNC are relatively well understood \cite{Adler-CoexistenceOf2-1990, Chesson-Generaltheoryof-2000, Muenkemueller-HutchinsonrevisitedPatterns-2009}, but robust tests for RNC in empirical studies are still scarce \cite{Descamps-Julien-StableCoexistencein-2005}. This may be because RNC is indeed rare in real communities, but equally plausible explanations are that RNC is comparatively difficult to detect \cite{Descamps-Julien-StableCoexistencein-2005}, or that empirical tests have concentrated on systems in which RNC is unlikely to occur \cite{Abrams-Whendoesperiodic-2004}. In particular, although a number of studies have linked resource fluctuations to coexistence, it requires fairly specific investigations to determine whether this link is mediated by relative nonlinearity, the temporal storage effect, or a simple niche-based mechanism in a stochastic environment \cite{Chesson-Quantifyingandtesting-2003, Descamps-Julien-StableCoexistencein-2005}. 

To examine the evolutionary properties and plausibility of RNC, we consider an evolutionary trade-off through which species have the option to invest into higher growth rates at high resource availability, at the expense of lower growth rates (and thus potentially even population crashes) at low resource availability. We describe this trade-off in terms of a density-compensation parameter in time-discrete population models with non-overlapping generations and several alternative density-dependence terms that follow classical population models (the Maynard Smith and Slatkin (MSS) model \cite{Smith-StabilityofPredator-Prey-1973}, the generalized Ricker model \cite{Ricker-Stockandrecruitment-1954}, and the Hassell model \cite{Hassell-Density-dependenceinsingle-species-1975}). Ecologically, this trade-off may be interpreted as representing how individuals use and monopolize available resources: resource monopolization strategies, such as scramble competition versus contest competition, or spatial resource distribution and searching behavior, for example, affect whether a population's growth reacts rather "weakly" (undercompensation), "normally" (compensation), or "strongly" (overcompensation) when its current size deviates from its carrying capacity \cite{Hassell-Density-dependenceinsingle-species-1975, Getz-HypothesisRegardingAbruptness-1996, Johst-Fromindividualinteractions-2008}. 

A previous study has shown that these model structures allow stable coexistence by RNC \cite{Muenkemueller-HutchinsonrevisitedPatterns-2009}. This earlier study, however, focused on community dynamics and did not provide an evolutionary analysis. Other studies did examine the evolutionary dynamics of parameters in the MSS model or similar models \cite{Felsenstein-randK-1979, Turelli-Densitydependentselection-1980, Doebeli-EvolutionofSimple-1995, Johst-Evolutionofcomplex-1999}. However, even though some of these studies also reported the existence of the aforementioned protected polymorphisms, none of them examined their evolutionary stability in detail (an exception is \cite{Metcalf-Evolutionofflowering-2008}, to which we relate our results in the Discussion). Moreover, some previous evolutionary studies exclusively relied on analytical investigations using adaptive dynamics theory and did not account for phenomena such as complex polymorphisms or demographic stochasticity, which are more easily captured through individual-based simulations. 

In this study, we address all these challenges together, to gain a more comprehensive appreciation of the role of relative nonlinearity for the evolution and maintenance of ecological diversity. We use adaptive dynamics theory and individual-based simulations to examine a number of variants of the assumed trade-off between a species' population growth rates at high and low resource availability. Our results allow us to draw conclusions about the ecological and evolutionary robustness of RNC as a coexistence mechanism, and highlight the need for combining ecological and evolutionary perspectives for understanding the process of community assembly and the emergence of biogeographic patterns.

\section*{Material and methods}

%
%

\subsection*{Dynamic vs. evolutionary stability of a coexistence mechanism}

To explain the methods of this paper, it will be useful to begin with some definitions and discuss how relative nonlinearity of competition maintains coexistence. In \cite{Chesson-Generaltheoryof-2000}, Chesson suggested to divide coexistence mechanisms into two classes: equalizing and stabilizing. Equalizing coexistence mechanisms reduce the fitness difference between species, and thus the speed of competitive exclusion. Stabilizing mechanisms, on the other hand, increase a species' fitness when its relative abundance (density) decreases, which actively stabilizes coexistence because it aids species when they become rare within a community. We refer to this type of stability as "dynamic stability", because stabilization acts on population dynamics on ecological time scales, as opposed to evolutionary stability, which refers to a stabilization of evolving genes or traits on evolutionary time scales. Since equalizing mechanisms make species effectively more "neutral", they must be expected to lead to evolutionary patterns and diversification processes similar to those predicted by neutral theories \cite{Chave-Neutraltheoryand-2004}. Stabilizing mechanisms, on the other hand, may be seen as generalizing the concept of the classical niche, because they increase the fitness of species at low densities. One might naively expect that stabilizing mechanisms will therefore also actively promote evolution towards species with such different, coexisting strategies. However, as confirmed by this study, the fact that a mechanism can dynamically stabilize coexistence is by no means a guarantee that selection will favor traits that create this stabilizing effect. 

The most straightforward mechanism to create dynamically stable coexistence in a non-spatial setting is based on the assumption that species use different resources. This leads to increasing fitness with decreasing frequency in a community, because species compete more strongly with their conspecifics. However, there are a number of further mechanisms that allow stable coexistence, even if those species use exactly the same resources. Those mechanisms include positive and negative interactions, such as facilitation or density-dependent mortality, the temporal storage effect, and relative nonlinearity of competition (RNC). Both RNC and the temporal storage effect are fluctuation-dependent mechanisms, meaning that they require non-constant resource availability over time. The difference between the two is the way coexistence is stabilized. The temporal storage effect is essentially caused by temporal niche partitioning, meaning that species have specialized on particular resource conditions that appear and disappear over time due to exogenously created resource dynamics. In RNC, on the other hand, species create or control resource fluctuations endogenously, and stabilization is being achieved because species affect fluctuations in a way that they limit their own fitness more than the fitness of their competitors.

\begin{figure*}[]
\centering
\includegraphics [width=13.5cm]{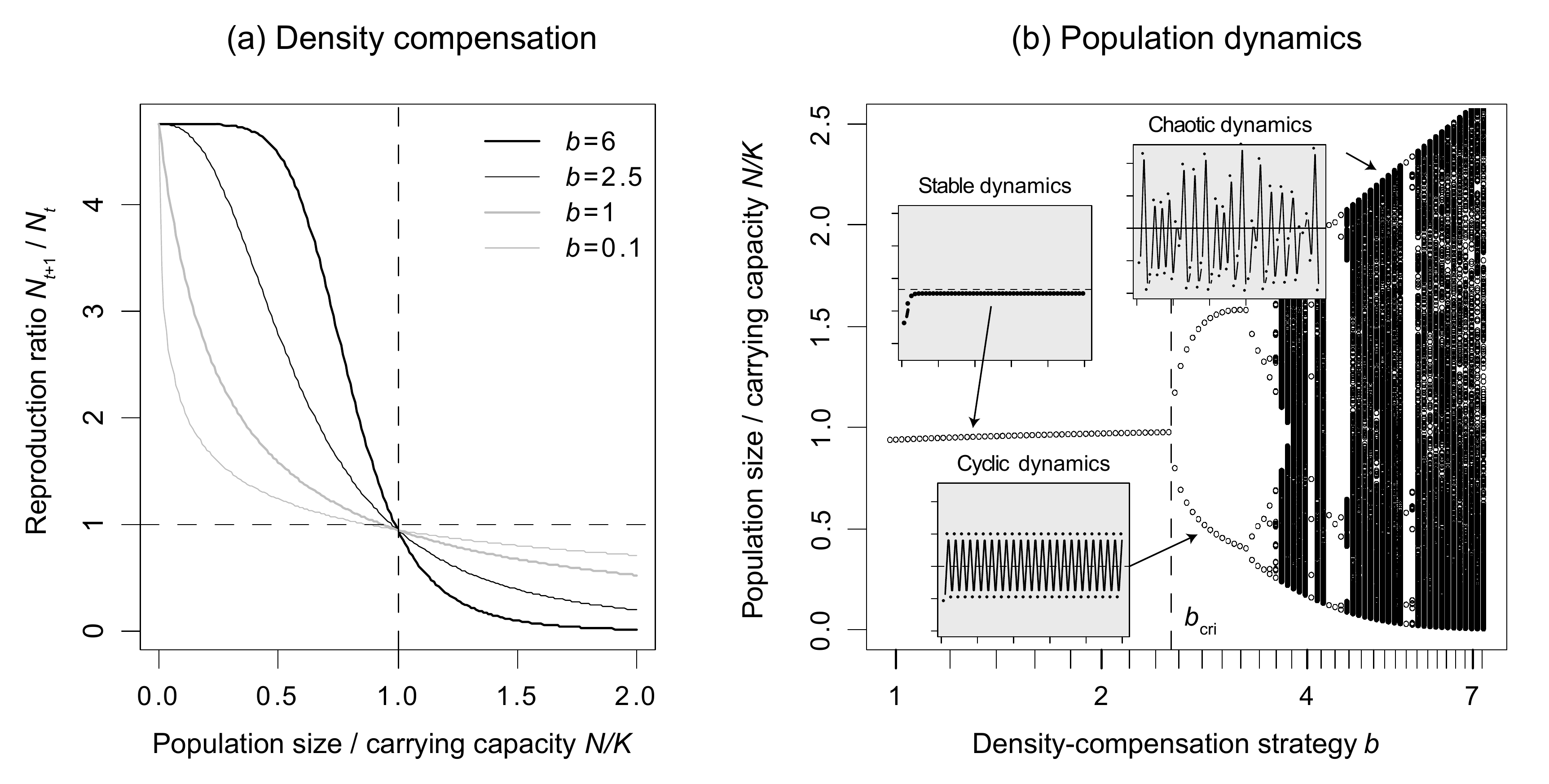} 
\caption{Effect of the density-compensation strategy $b$ on the population dynamics of a single species described by the Maynard Smith and Slatkin model. (a) Reproduction ratio $N_{t+1} / N_t$ as a function of the relative population size $N/K$ for the analytical MSS model with  intrinsic growth rate $r=5$, density-independent mortality $d=0.05$, and four different values of $b$. Since $d>0$, the reproduction ratio at the carrying capacity $K$ is smaller than $1$, which implies that the equilibrium population size remains slightly below the carrying capacity. (b) Bifurcation diagram showing population sizes at equilibrium as a function of the density-compensation strategy $b$. Cyclic population dynamics, indicating strong overcompensation, occur for $b$-values exceeding $ b_{\mathrm{cri}}\approx 2.5 $ (vertical line), as predicted by eq.~\ref{eq: criticalvalue for MSS}. The three insets depict the transition from stable population dynamics below the critical value to cyclic dynamics shortly above the critical value and to chaotic dynamics at even larger $b$-values. \label{figure: growth curves}} 
\end{figure*}

\subsection*{RNC in models with nonlinear density dependence}

Relative nonlinearity has frequently been studied using models that explicitly describe resource and consumer dynamics \cite{Armstrong-CompetitiveExclusion-1980}, but stabilization by RNC is also possible in simple time-discrete models of density-dependent population growth, in which resource availability is implicitly described by a shared carrying capacity \cite{Muenkemueller-HutchinsonrevisitedPatterns-2009}. This corresponds to resources, such as light or space, for which the overall resource availability is constant, but it should also be a good approximation for situations where resource dynamics are fast compared to population dynamics. 

We describe the reproduction of a population with non-overlapping generations from population size $N$ at time $t$ to size $N_{t+1}$ at time $t+1$ by density-dependent growth rates derived from three widely used models (the Maynard Smith and Slatkin (MSS) model \cite{Smith-StabilityofPredator-Prey-1973}, the generalized Ricker model \cite{Ricker-Stockandrecruitment-1954}, and the Hassell model \cite{Hassell-Density-dependenceinsingle-species-1975}). In each model, the reproduction ratio $f(N_t) = N_{t+1}/N_t$ depends on the population size (i.e. the population density) $N$, on the intrinsic growth rate $r$, on the carrying capacity $K$, and on a parameter $b$ that controls the shape of the density compensation (Fig.~\ref{figure: growth curves}). In all three models, we included an additional parameter $d$ that was not present in the original model equations. This parameter describes a density-independent mortality risk of individuals, which may originate, for example, from external disturbances. The motivation for including such a term will be discussed later. To distinguish the population-level models from the individual-based models described later, we refer to the following eq.~\ref{eq: MSequation} as the analytical MSS model, and to the other models accordingly. 

The analytical MSS model is given by
\begin{equation}\label{eq: MSequation}
    f_\mathrm{MSS}(N) =  \frac{(1-d) \cdot r}{ 1+(r-1) \cdot(N/K)^{b}}\;.
\end{equation}
The functional form of $f_\mathrm{MSS}(N)$ for different values of $b$ is displayed in Fig.~\ref{figure: growth curves}. The analytical generalized Ricker model is given by 
\begin{equation}
f_\mathrm{Ricker}(N)= (1-d) \cdot e^{r \cdot [1- (N /K)^b]}\;.
\label{eq: Ricker}
\end{equation}
Here, the term "generalized" refers to the exponent $b$ in the equation above, which, for $b \neq 1$, provides an extension of the classical Ricker model \cite{Ricker-Stockandrecruitment-1954}. The analytical Hassell model is given by
\begin{equation}
f_\mathrm{Hassell}(N)= \frac{(1-d) \cdot r }{[ 1 + (r^{1/b}-1) \cdot N/K]^{b}}\;.
\label{eq: Hassel}
\end{equation}
The term $(r^{1/b}-1)$ in the denominator is a common reformulation of the Hassell model. It allows translating the parameter $a$ used in the original version of this model \cite{Hassell-Density-dependenceinsingle-species-1975} into a carrying capacity $K$, which makes the model parameters more comparable to those of the Ricker model and the MSS model.

\subsection*{Overcompensation creates population fluctuations}

It is well known that eqs.~\ref{eq: MSequation},\ref{eq: Ricker},\ref{eq: Hassel} may produce cyclic or chaotic population dynamics, depending on the values of $r, b$ and $d$. For our further analysis, it will be useful to determine the critical value $b_\mathrm{cri}$ at which the population dynamics start to exhibit cycles. Oscillations start when a deviation of the population size from its equilibrium leads to a compensation that is stronger than the original deviation (overcompensation). This motivates the definition of the complexity $c$ as the derivative of the population-level reproduction $f(N) \cdot N$ with respect to $N$, evaluated at the equilibrium population size $N^\star$, which is defined by $f(N^\star) = 1$:
\begin{equation}\label{eq: definition complexity}
	c = \mathrm{d}/\mathrm{d}N (f(N) \cdot N) |_{N=N^\star} \;.
\end{equation}
If $c < -1 $, a deviation from the equilibrium is compensated by an even larger deviation in the opposite direction. With $c = -1$, solving eq.~\ref{eq: definition complexity} for $b$ yields the critical value $b_\mathrm{cri}$ as a function of $r$ and $d$. For eq.~\ref{eq: MSequation}, the result is
\begin{equation}\label{eq: criticalvalue for MSS}
	b_\mathrm{cri}(r,d) = 2 - \frac{2}{1 + (d-1) \cdot r} \;.
\end{equation}
The population dynamics for different values of $b$ in eq.~\ref{eq: MSequation} are illustrated in Fig.~\ref{figure: growth curves}. The critical values for the Ricker and the Hassell model are determined analogously. 

\subsection*{Adaptive dynamics for analyzing evolutionary and dynamic stability}

To examine coexistence in the models defined above, we consider two species reproducing according to eqs.~\ref{eq: MSequation},\ref{eq: Ricker},\ref{eq: Hassel} that share the same resources, but differ in their $b$-strategy. For the MSS model, this results in the following species-specific reproduction ratios,

\begin{equation}\label{eq: MSequation-2}
\begin{split}
  f_1(N_1,N_2) & =  \frac{(1-d) \cdot r}{ 1+(r-1) \cdot((N_1+N_2)/K)^{b_1}}  \;, \\
  f_2(N_1,N_2) & =  \frac{(1-d) \cdot r}{ 1+(r-1) \cdot((N_1+N_2)/K)^{b_2}} \;,
\end{split}
\end{equation}
where the fact that the species use the same resource is evident because the reproduction ratios depend on the sum $N = N_1 + N_2$ of the population sizes of the individual species. We treat cases with more species and other growth models accordingly. 

To assess dynamic and evolutionary stability for the coupled system given by eq.~\ref{eq: MSequation-2}, we use pairwise invasibility plots. These plots show the fitness $f$ of a rare mutant that attempts to invade a resident community at equilibrium population size ($N_\mathrm{I} \ll K, N_\mathrm{R} \approx K$). We follow the standard definition of invasion fitness $f$ as the invader's average (natural) logarithmic growth rate during a large number of $T$ generations 

\begin{equation} \label{eq: invasion fitness}
	f = \frac{1}{T}\sum_{t=1}^T \mathrm{log}\frac{N_{\mathrm{I},t}}{N_{\mathrm{I},t-1}}.
\end{equation}

Averaging over $T$ is necessary to account for resident populations with cyclic or chaotic population dynamics. We chose $T=500$ throughout this study to obtain a representative sample of resident population sizes even in the chaotic regime. 

The shape of the pairwise invasibility plots allows a visual assessment of dynamic stabilization and the probable evolutionary dynamics of an evolving strategy. Mutual invasibility of two strategies, for example, indicates dynamic stability, as there is a fitness advantages for both species when they are at low relative frequency. A discussion of how to interpret such plots with regard to evolutionary dynamics and evolutionary stability can be found in \cite{Dieckmann-Canadaptivedynamics-1997}.

\subsection*{Trade-offs between growth rates at low and high resource availability}

For using the adaptive dynamics framework described above, we have to decide which of the species traits that are coded as parameters in eq.~\ref{eq: MSequation-2} are allowed to evolve. It is known that coexistence by RNC can arise in population models such as eqs.~\ref{eq: MSequation},\ref{eq: Ricker},\ref{eq: Hassel}; and if it does, it occurs between a species favored at high resource availability, and a species favored at low resource availability \cite{Muenkemueller-HutchinsonrevisitedPatterns-2009}. We thus consider a trade-off between growth rates at high and low resource availability. This is ecologically plausible, and can be mechanistically motivated by the various ways in which species utilize and monopolize their available resources (e.g. in the contexts of contest versus scramble competition or of spatial distribution patterns; see also \cite{Hassell-Density-dependenceinsingle-species-1975, Getz-HypothesisRegardingAbruptness-1996, Johst-Fromindividualinteractions-2008}). 

A convenient way to create families of density-dependence functions that respect this trade-off is to vary the parameter $b$ in eqs.~\ref{eq: MSequation},\ref{eq: Ricker},\ref{eq: Hassel}. As can be seen in Fig.~\ref{figure: growth curves}, increasing $b$ leads to higher growth at population sizes below the carrying capacity $K$, and to lower growth otherwise. There would certainly also be other ways to create families of density-dependence functions that respect such a trade-off. For example, one could consider varying $r$ as well. However, varying $r$ mostly affects growth at low population densities, and varying both $b$ and $r$ without further constraints is not possible, because the single best option for a species is then to have a large $r$ and a small $b$, which results in comparably favorable growth rates both above and below the carrying capacity \cite{Doebeli-EvolutionofSimple-1995}. Thus, varying $b$ in the three models is the most straightforward option for creating ecologically reasonable and smoothly changing families of density-dependence functions. 

It is beyond the scope of this paper to analyze all possible further families of curves that respect the trade-off described above, but we will examine a particular modification of the Maynard Smith and Slatkin (MSS) model later, to further explore the generality of our conclusions. As the motivation for this modification originates from our results, we provide the specification and further explanation of this modification as part of the Results. 

The aim of creating a trade-off between fitness at high and low resource availability is also the motivation for introducing the density-independent mortality $d$ that was described earlier. This mortality is not part of the original models, but without $d$, all subcritical density-dependence functions (i.e. those leading to stable population dynamics) would result in equilibrium population sizes exactly matching the carrying capacity  $K$, where all growth rates are identical, regardless of the value of $b$. As a result, all those strategies would be subject to neutral drift. By introducing $d$, changes in $b$ always lead to effective fitness differences and therefore to a real trade-off, and not just equal fitness, in $b$.

\subsection*{Individual-based simulations of evolution and coexistence}

To test whether our results based on adaptive dynamics theory are robust under demographic stochasticity and when allowing for more complex polymorphic strategies, we repeat parts of the analysis by explicitly simulating the evolutionary dynamics with an individual-based model (IBM). To maintain comparability, the IBM implements exactly the same ecological processes that we considered in the analytical models, except that reproduction is now stochastic and that it is possible to model as many strategies $b_i$ as there are individuals $i = 1, \ldots N $. Because the adaptive dynamics analysis revealed the absence of qualitative differences between eqs.~\ref{eq: MSequation},\ref{eq: Ricker},\ref{eq: Hassel} with respect to the key questions addressed in this paper, we restrict the presentation of IBM results to the MSS model. 

In the MSS IBM, we assume that the reproduction of individuals shows the same density dependence as in the analytical model introduced earlier. Thus, an individual $i$ produces offspring $n_i$ according to the MSS density dependence,

\begin{equation}\label{eq: MSequation for individuals}
    n_{i, t+1} \sim  \ P [ f(N_t/K; b_i, r, d) ].
\end{equation} 

with the MSS reproduction ratio $f(N_t/K; b_i, r, d)$ from eq.~\ref{eq: MSequation} depending on the total population size $N_t$ (sum of all individuals) divided by the population carrying capacity $K$. The two main differences to the analytical model are that the IBM allows a different density-compensation strategy $b_i$ for each individual, and that the number of offspring is drawn from a Poisson distribution, indicated by $P$. The latter ensures that each individual produces an integer number of offspring, and that demographic stochasticity, which is present in all natural populations, is accounted for, including the possible extinctions of strategies. For a large number of individuals and all $b_i$ being equal, the IBM recovers the analytical model eq.~\ref{eq: MSequation}. Simulation code for the IBM and for the analytical models is available at \href{https://github.com/florianhartig/EvolutionOfRelativeNonlinearity/}{https://github.com/florianhartig/EvolutionOfRelativeNonlinearity/}.

Because strategies can go extinct in the IBM, especially during invasion, we use a different measure of invasion fitness than for the analytical models (eq.~\ref{eq: invasion fitness}), namely the probability $p$ that a strategy invading with one individual survives for at least 500 generations in a resident population at equilibrium. For computational efficiency, we approximate this value by simulating invasions with three individuals, resulting in estimates of the probability $q^{(3)}$ that a mutant strategy invading with three initial individuals goes extinct after 500 generations. This probability then yields the probability $q = 1-p$ that one individual goes extinct according to $q = \sqrt[3]{q^{(3)}}$ (assuming that most extinctions due to demographic stochasticity occur soon after the invasion, so that the three individuals are approximately independent). Additionally, we examine coexistence times (defined as the expected times to competitive exclusion starting from equal population sizes) for different pairs of density-compensation strategies, which allows to determine which strategy is outcompeted in the long run.

\begin{figure*}[]
\centering
\includegraphics [width=17.35cm]{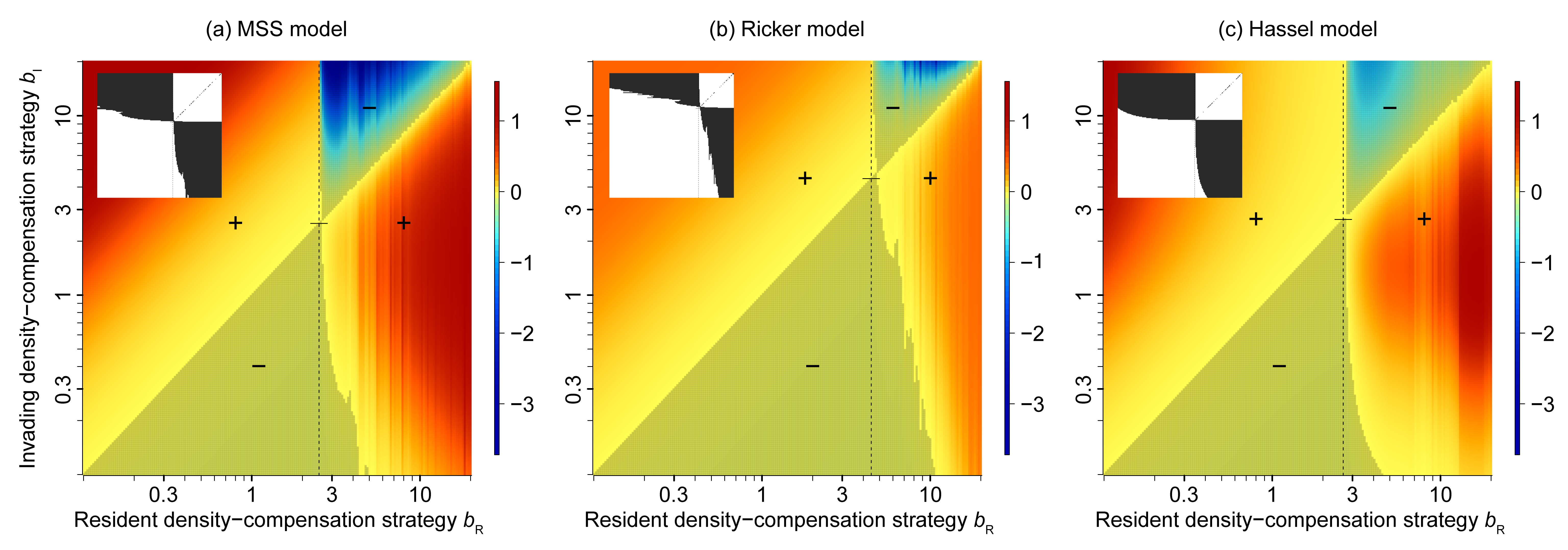} 
\caption{Pairwise invasibility plots for different density-compensation strategies $b$ in eqs.~\ref{eq: MSequation},\ref{eq: Ricker},\ref{eq: Hassel}. The plots show the fitness (eq.~\ref{eq: invasion fitness}) of an invading strategy with a density-compensation strategy along the vertical axis for residents with density-compensation strategies along the horizontal axis. Plus and minus signs indicate strategy combinations resulting in positive and negative invasion fitness, respectively; in addition, regions of negative invasion fitness are shaded. Vertical dashed lines show the critical $b$-values of the resident. Small insets in the top left of each plot show areas of mutual invasibility. Parameters: $r=5$ (MSS),  $r=40$ (Hassell), $r=0.5$ (Ricker), and $d=0.05$. Different intrinsic growth rates $r$ were chosen to obtain a similar growth response for similar $b$-values across the three models. \label{figure: classical PIP}} 
\end{figure*}

To include evolution in the IBM, we implement mutations of $b$-values that occur, for each individual, with a probability $m / K$ before reproduction. The division by $K$ is introduced to facilitate the interpretation of $m$: for $m=1$, there is on average one mutation per generation when the population is at its carrying capacity. If a mutation occurs, a normal random variable with zero mean and standard deviation $\sigma$ is added to the parental strategy. Individuals with different $b$-values produce different numbers of offspring at different population sizes. This creates selection on the density-compensation strategy $b$, and thereby drives evolution in response to the experienced environmental conditions. Ecologically, it does not make sense for the evolving strategy $b$ to get too close to $0$, because this would make the reproduction independent of the population size. Therefore, we introduce a cutoff parameter $b_\mathrm{min}=0.17$. Mutations with $b<b_\mathrm{min}$ are set to $b_\mathrm{min}$. This value of $b_\mathrm{min}=0.17$ is considerably lower than any $b$-values that will be important for the analysis. Hence, the introduction of $b_\mathrm{min}$ is only a technical safeguard and has no influence on our results.

\begin{figure*}[]
\centering
\includegraphics [width=13.5cm]{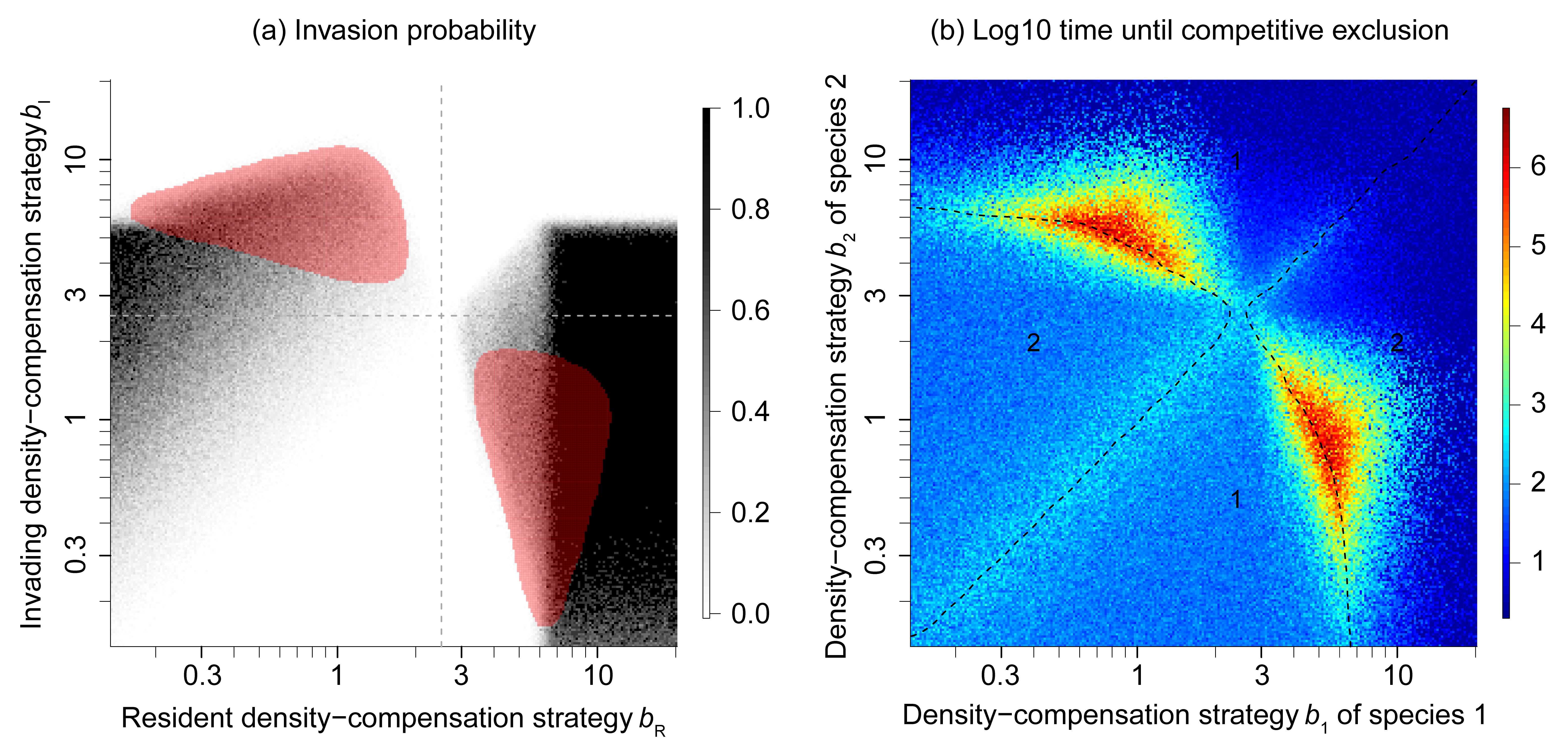} 
\caption{Evolutionary and dynamic stability in the individual-based Maynard Smith and Slatkin model. (a) Invasion probability, approximated by the probability that a strategy invading with one individual survives for at least 500 generations. Red shades depict areas that are mutually invasible. Dashed lines indicate the critical density-compensation strategy at $b_{\mathrm{cri}} \approx 2.5$ (eq.~\ref{eq: criticalvalue for MSS}). (b) Time until competitive exclusion (plotted in log10 units) for two species, starting with equal population sizes. The dashed curves (obtained using a kernel smoother) show combinations of $b$-values for which the two strategies have equal chances to exclude each other. The diagonal line (identical $b$-values) may be regarded as providing a reference: it shows the time until competitive exclusion under neutral drift. Moving away from the diagonal, one of the two strategies (marked by the numbers 1 and 2) tends to exclude the other, with an average time until competitive exclusion smaller than under neutral drift. More interesting, however, are the coexistence times along the other parts of the dashed curves, which are several orders of magnitude longer than along the diagonal, evidencing a non-neutral, stabilizing mechanism of coexistence. Each cell shows the results from a single simulation; hence, the variance among close-by cells provides a visual impression of the variance between simulation runs. Other parameters: $K=200$, $r=5$, and $d=0.05$. \label{figure: coexistence}} 
\end{figure*}


To examine the consequences of introducing evolution, we first test how a community with previously fixed density-compensation strategies is affected by the possibility of mutations in $b$. Then, we calculate the $b$-strategy that is attained by evolution in the long run. To record the evolutionary equilibrium, we allow $10^6$ generations for convergence before data acquisition is started. As we find no path dependence in the IBM, we eschew replicate models runs for the same parameter values in favor of a finer coverage of the parameter space: the local fluctuations in the results then allow a visual impression of the variability among model runs. Simulations were initialized with $N=K$ individuals, each of which is assigned a different density-compensation strategy $b$ drawn from a uniform distribution in the interval $[b_\mathrm{min}, 15]$. Initialization with such a random ensemble enables faster convergence to the evolutionarily equilibrium than starting from a single strategy. 

Unlike for the analytical models, where dynamics are not dependent on the carrying capacity $K$ apart from trivial rescaling, changing $K$ in the IBM may affect demographic stochasticity and genetic drift. The smaller $K$, the larger the relative strength of population fluctuations created by demographic stochasticity, thus increasing the speed of genetic drift. Consequently, the outcomes of evolution in the density-compensation strategy $b$ differ most between the analytical model and the IBM when the carrying capacity $K$ is small. As we aim to assess whether these differences may affect the evolutionary dynamics, it is important that $K$ is not chosen too large (we use values of $K=200$ and $K=1000$). Apart from that, however, there are no indications that the specific choice of $K$ qualitatively affects our conclusions regarding the stability of coexistence. For these reasons, we do not systematically examine the effect of varying $K$.

\section*{Results}

\subsection*{Dynamics predicted from adaptive dynamics theory}

The first part of our results uses the analytical population models eqs.~\ref{eq: MSequation},\ref{eq: Ricker},\ref{eq: Hassel} to examine how the fitness of an invading density-compensation strategy depends on the density-compensation strategy of the resident population. From the resulting pairwise invasibility plots, one can deduce the strategy, or strategy combinations, that are dynamically or evolutionarily stable \cite{Dieckmann-Canadaptivedynamics-1997}.

A first observation to highlight is that invasibility patterns change at an intermediate $b$-value of the resident population (Fig.~\ref{figure: classical PIP}). Numerical calculations (eq.~\ref{eq: definition complexity}) confirm that, for all models, this $b$-value coincides with the critical value at which resident population dynamics start to exhibit cycles and, for larger $b$-values, chaotic dynamics. Resident populations below the critical $b$-value ($b < b_{\mathrm{cri}}$), that is, residents with stable population dynamics, can generally be invaded by strategies with stronger density compensation (larger $b$). The reason is that the small value of $d=0.05$ chosen here is sufficient to ensure that subcritical $b$-values result in equilibrium population sizes slightly below the carrying capacity, which favors higher $b$-values. Resident populations with cyclic or chaotic dynamics ($b > b_\mathrm{cri}$), on the other hand, can generally be invaded by strategies with weaker density compensation (smaller $b$) than the resident. There is a fairly broad range of $b$-values that are mutually invasible, which indicates dynamically stable coexistence of the corresponding strategy pairs. In Fig.~\ref{figure: classical PIP}, we highlight those values in the insets at the top left corner of each panel. 

A further analysis of the pairwise invasibility plots, however, suggests that these pairs of mutually invasible strategies are not evolutionarily stable. There exists one intermediate strategy, slightly above $b_\mathrm{cri}$, that cannot be invaded by any other strategy. The shape of the invasion fitness around this so-called evolutionarily singular point suggests that it is the only evolutionarily stable strategy \cite{Dieckmann-Canadaptivedynamics-1997}. Below, we will confirm this conclusion with the IBM.

\subsection*{Invasibility and coexistence in the IBM}\label{sec: pairwise invasibility}

The pairwise invasibility plots in Fig.~\ref{figure: classical PIP} allow deducing the probable evolutionary dynamics, but are limited in that they consider only strategy pairs that reproduce deterministically. Complex polymorphic strategies, demographic stochasticity and effects of small population sizes, as well as strategy extinctions are not considered in such an analysis. For this reason, we repeat the analysis of evolutionary and dynamic stability of RNC using the individual-based MSS model. 

The resulting pairwise invasibility plot (Fig.~\ref{figure: coexistence}a) shows a similar pattern as Fig.~\ref{figure: classical PIP}a. However, in contrast to the analytical model, strategies with large $b$-values can  generally not invade. The reason is not that they produce too little growth, but rather that these strategies are generally not viable, because they imply population fluctuations that are so strong that they quickly drive the population to extinction (note that strategies cannot go extinct in the analytical models). One can think of invasibility in the IBM as resulting from two requirements: positive growth, and dynamic persistence of the population. Still, there are relatively large ranges of $b$-strategies that are mutually invasible (red shaded areas in Fig.~\ref{figure: coexistence}a), indicating dynamic stability. As for the analytical model, we find that one particular intermediate density-compensation strategy cannot be invaded by any other strategy. The shape of the pairwise invasibility plot around this singular strategy again suggests that it is evolutionarily stable \cite{Dieckmann-Canadaptivedynamics-1997}. The $b$-value of this strategy approximately corresponds to $b_\mathrm{cri} $, meaning that the evolutionarily stable strategy (ESS) is located where the population dynamics start to exhibit cycles.

\begin{figure*}
\centering
\includegraphics [width=13.5cm]{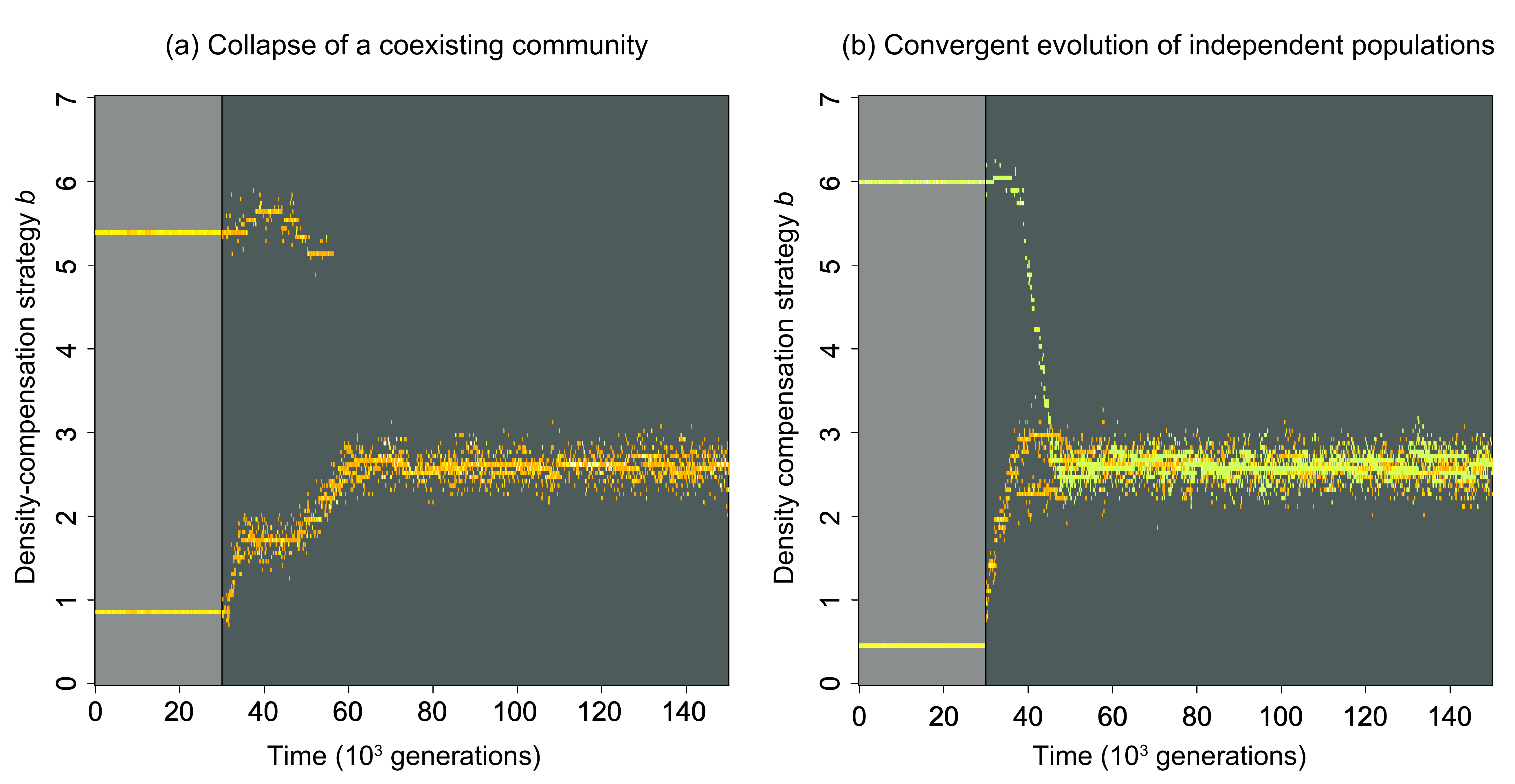} 
\caption{Evolutionary convergence to a single density-compensation strategy in the individual-based Maynard Smith and Slatkin model. In both panels, evolution is absent for the first $3\cdot 10^5$ generations (distinguished by a lighter background) and present thereafter. (a) Evolution of the density-compensation strategies of two coexisting species, starting from initial values that are known from Fig.~\ref{figure: coexistence} to enable dynamically stable coexistence ($b_1=0.9,b_2 =5.4$). After evolution starts, the species rapidly evolve outside the coexistence region, which leads to the extinction of one species and the evolution of the other to an intermediate density-compensation strategy. (b) Evolution of the density-compensation strategies of two isolated (non-coexisting) species, starting from two different $b$-values ($b_1 = 0.5, b_2 = 6$). After evolution starts, the species rapidly evolve to the same $b$-value as in (a). Other parameters: $K = 1000$, $\sigma=0.15$, $m = 0.3$, $r=5$, and $d=0.05$.\label{figure: collapse}} 
\end{figure*}

Examining the dynamic stability of strategy pairs in terms of their average time to competitive exclusion (Fig.~\ref{figure: coexistence}b) reveals the strength of the stabilization inferred from the mutual invasibility in the analytical models: in addition to the "neutral" pairs along the diagonal, where both strategies are trivially of equal fitness, there is a second curve of strategies that have equal fitness, but consist of one species with a weak and one species with a strong density-compensation strategy. This second curve overlaps with the mutually invasible areas found in Fig.~\ref{figure: coexistence}a. Within those areas, we find strategy pairs with very different $b$-values that allow coexistence times up to four orders of magnitude longer than strategy pairs along the diagonal. This shows that the former strategies are not simply coexisting neutrally, but that their coexistence is actively stabilized, confirming what we conjectured based on the mutual invasibility results discussed in the previous paragraph. The underlying mechanism is that an overcompensating (higher) $b$-strategy has an advantage within a predominantly undercompensating population as long as the total population size is below the carrying capacity. The larger the relative frequency of the overcompensating strategy, however, the higher the probability that the population is overshooting its carrying capacity. At those times, the undercompensating strategy is advantageous. Thus, neither species can outcompete the other, because each of them creates an advantage for the other one as soon as it becomes dominating (a detailed analysis of those dynamics is provided by \cite{Muenkemueller-HutchinsonrevisitedPatterns-2009}).

\begin{figure*}[]
\centering
\includegraphics [width=13.5cm]{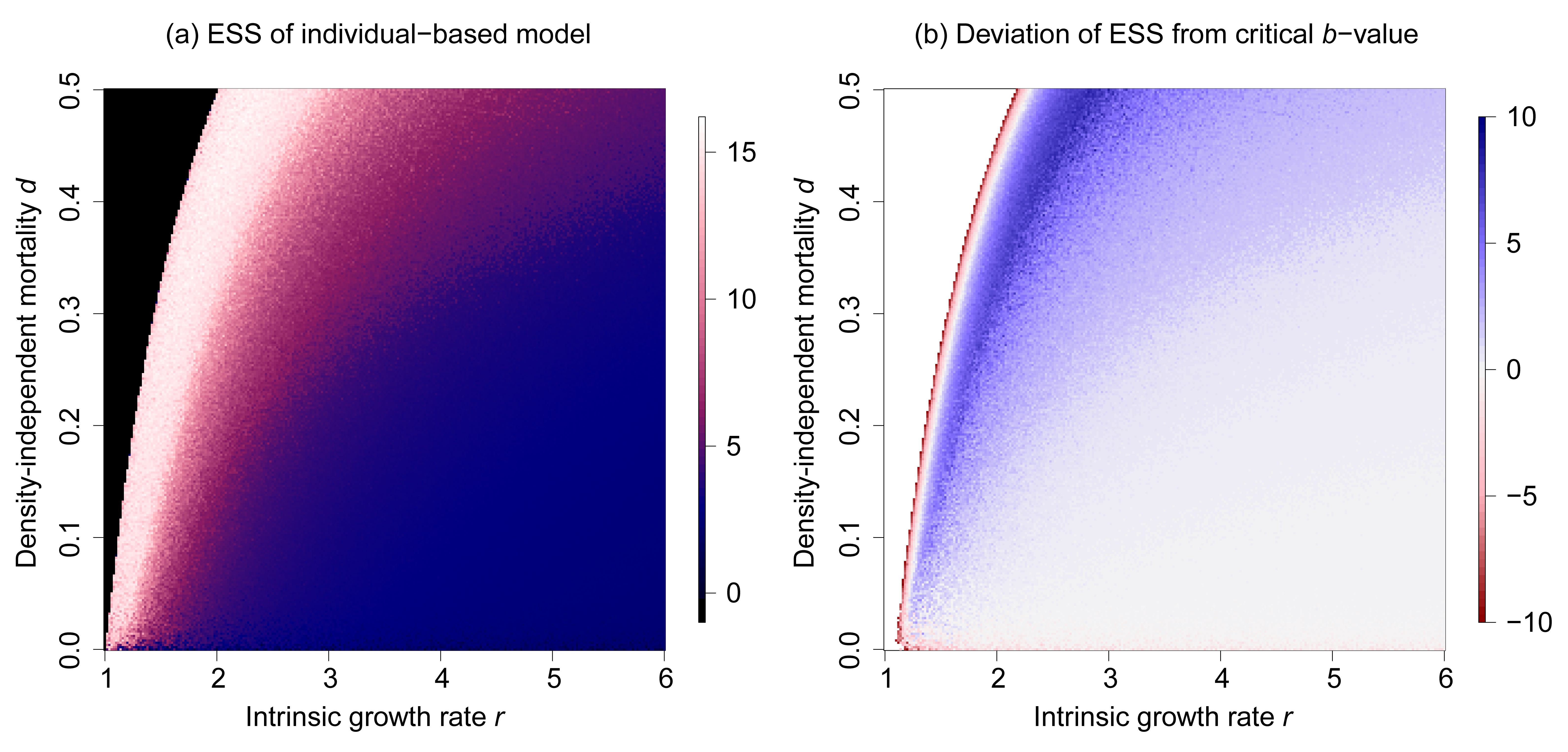} 
\caption{Evolutionarily stable density-compensation strategies in the individual-based Maynard Smith and Slatkin model. (a) Evolutionarily stable $b$-values as a function of the intrinsic growth rate $r$ and density-independent mortality $d$. Black colors indicate extinction. (b) Difference between the evolutionarily stable strategy and the critical $b$-value (eq.~\ref{eq: criticalvalue for MSS}). Other parameters: $K = 1000$, $\sigma=0.05$, $m = 0.1$, and $b_\mathrm{min} = 0.17$.\label{figure: localEvolution}} 
\end{figure*}

\begin{figure*}[]
\centering
\includegraphics [width=17.35cm]{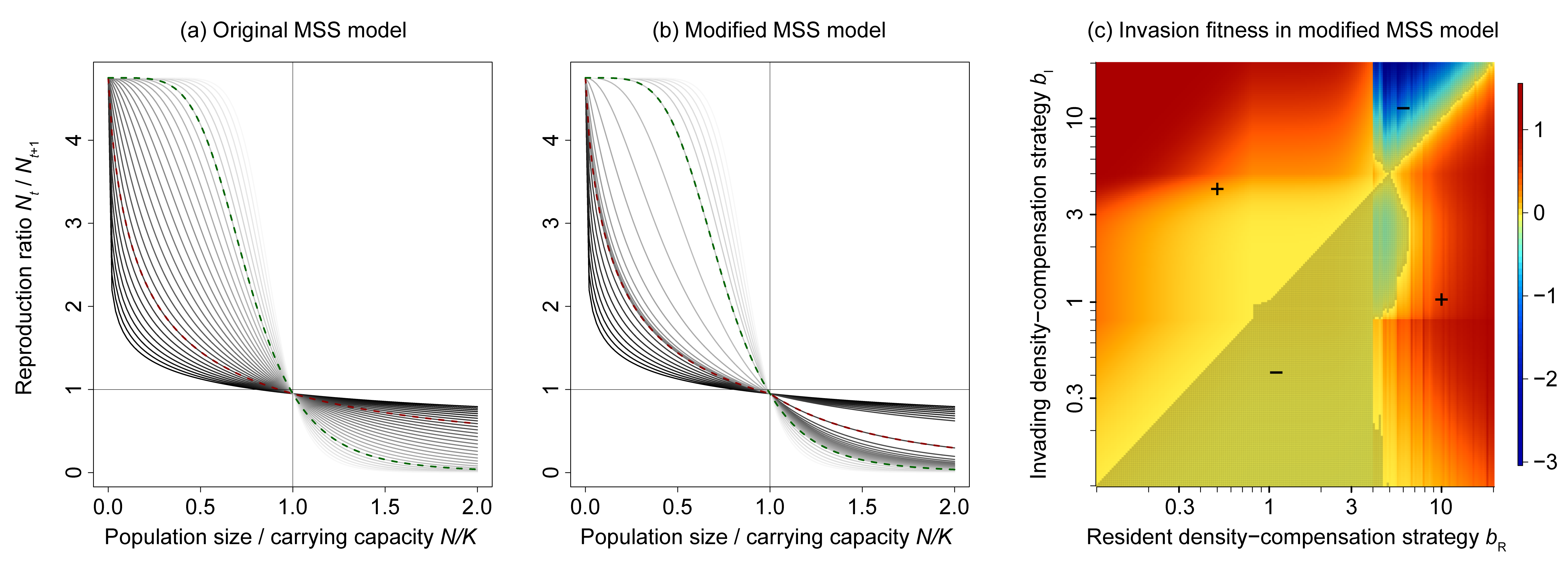} 
\caption{Density-dependent reproduction ratio of the original Maynard Smith and Slatkin model (a), of the modified Maynard Smith and Slatkin model (b), and resulting pairwise invasibility plot (c). The curves plotted in (a) and (b) result from equidistant $b$-values on the logarithmic scale. The curves for $b_\mathrm{low}=0.8$ (red) and $b_\mathrm{up}=5$ (green) from eq.~\ref{eq: modified MSS} are highlighted by dashed curves. Note that, while curves are evenly distributed for the original model, the modification creates an asymmetry between the curves above and below $N/K=1$ in the modified model. The pairwise invasibility plot shows that this results in lower fitness for the intermediate strategies, as well as in the loss of evolutionary stability of the evolutionarily singular strategy. \label{figure: pip-mod}} 
\end{figure*}

\subsection*{Evolutionary dynamics in the IBM}

Introducing evolution in the individual-based MSS model, by incorporating mutations as described in the Methods, confirms the results of the invasibility analysis: there is one evolutionary attractor close to the critical $b$-value, and the strategy associated with this attractor is attained irrespective of whether we start from a single density-compensation strategy (Fig.~\ref{figure: collapse}b), or from a community that coexists by RNC (Fig.~\ref{figure: collapse}a). The $b$-value of this strategy approximately coincides with the critical $b$-value derived from eq.~\ref{eq: criticalvalue for MSS}, and with the evolutionarily singular strategy discussed in the previous section (Fig.~\ref{figure: coexistence}a). This confirms that this singular strategy is not only globally evolutionarily stable, but also globally evolutionarily attainable. To test whether the same conclusion holds also for other choices of $r$ and $d$, we varied those parameters systematically. We find that evolution always leads to a unique, evolutionarily stable $b$-value, which generally seems to coincide approximately with the critical $b$-value at which the population dynamics start to exhibit cycles (Fig.~\ref{figure: localEvolution}). We conjecture that differences to the analytically expected singular points result from demographic stochasticity and different equilibrium population sizes, which depend on $r$ and $d$.

\begin{figure*}[]
\centering
\includegraphics [width=13.5cm]{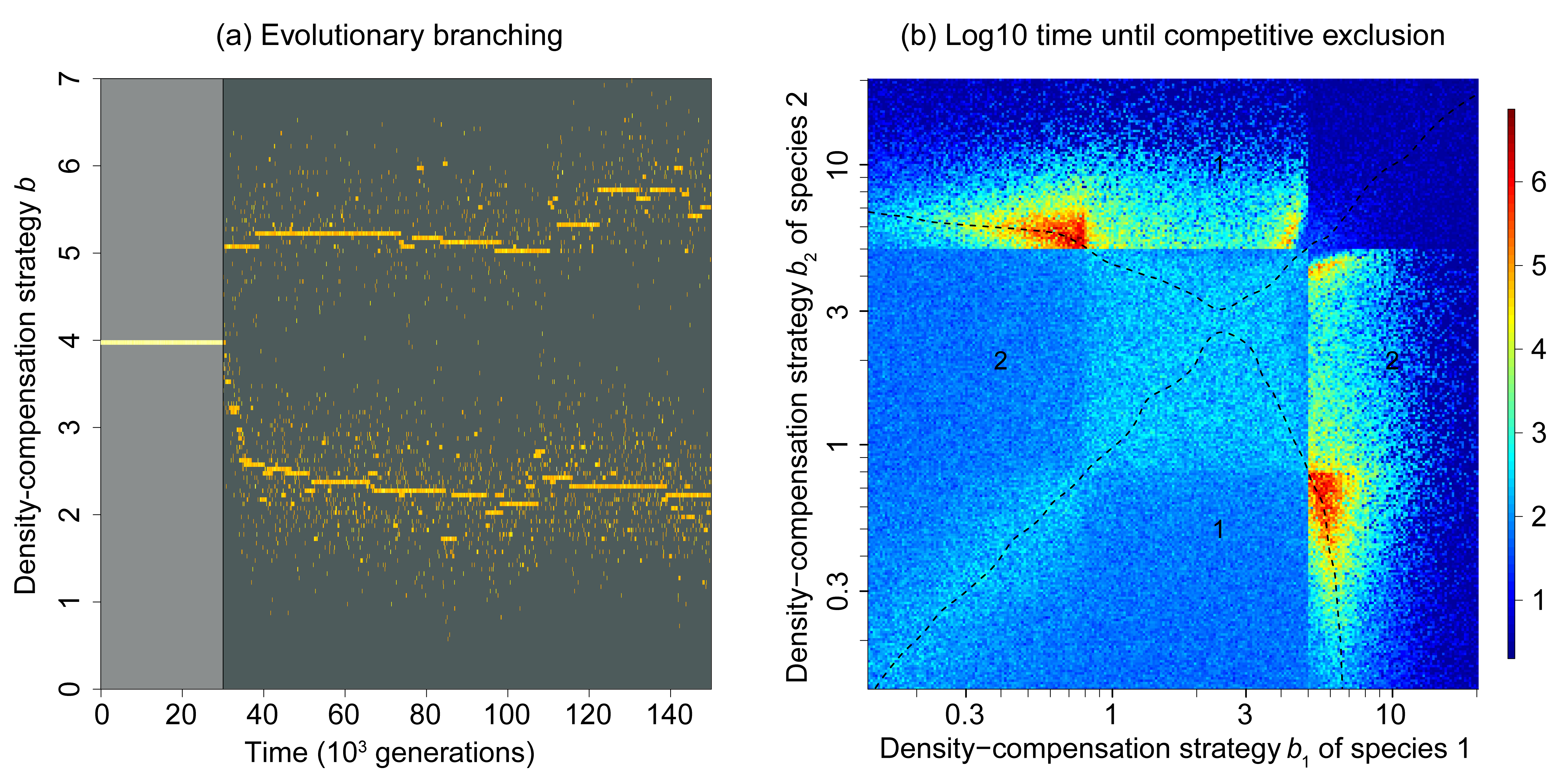} 
\caption{Evolutionary branching (a) and time until competitive exclusion (b) for the modified Maynard Smith and Slatkin model (eq.~\ref{eq: modified MSS}). After evolution is introduced (distinguished by a darker background), evolutionary branching results in two distinct strategies that are evolutionarily stable. Analysis of the time until competitive exclusion (plotted in log10 units) indicates that these strategies are also dynamically stabilized by RNC. Other parameters: $K = 1000$, $\sigma=0.3$, $m = 0.1$, and $b_\mathrm{min} = 0.17$.\label{figure: evolution-mod}} 
\end{figure*}

\subsection*{Creating evolutionarily stable RNC}

For all models investigated so far, we can conclude that a large number of $b$-strategy pairs can dynamically coexist, but none of these strategy pairs is evolutionarily stable. This raises the question whether the trade-off between growth at high resource availability and persistence at low resource availability can result in evolutionarily stable coexistence at all. To answer this questions, we systematically modified the MSS equation to reduce fitness in the neighborhood of the evolutionarily stable strategy. We changed $f_\mathrm{MSS}(N)$ (eq.~\ref{eq: MSequation}) so that for all $b_\mathrm{low} = 0.8 < b < b_\mathrm{up} = 5$, the $b$-value used in eq.~\ref{eq: MSequation} is replaced by a value $b_\mathrm{m}$ that depends on $N$ as follows, 

\begin{equation}\label{eq: modified MSS}
 b_\mathrm{m} =\begin{cases}
     b_{\mathrm{low}} + \frac{b^\tau }{ (b_\mathrm{up}-b_\mathrm{low})^{\tau-1}} & \text{if $N<K$},\\
     b_{\mathrm{low}} + \frac{b^{1/\tau} }{ (b_\mathrm{up}-b_\mathrm{low})^{1/\tau-1}} &\text{if $N>K$}.
  \end{cases}
\end{equation}

After the modification, the density-dependence functions that previously changed rather gradually with changes in $b$ above and below $N$ now change nonlinearly and with different speeds depending on whether $N$ is below or above the carrying capacity (Fig.~\ref{figure: pip-mod}). This nonlinear effect is controlled by a scaling parameter $\tau$ that we set to $\tau=4$. The interval in which this change happens, between $0.8<b<5$, is deliberately focused on the area between the $b$-values that resulted in stable coexistence in our previous analysis. 

As a result of this modification, the shape of the pairwise invasibility plot is changed (Fig.~\ref{figure: pip-mod}c). There no longer exists a strategy that is stable against invasion by any other strategy. Our individual-based simulations show that this indeed creates disruptive selection towards two distinct density-compensation strategies (Fig.~\ref{figure: evolution-mod}a). Looking at the time to competitive exclusion for these strategy pairs confirms that they are stabilized by RNC (Fig.~\ref{figure: evolution-mod}b), although evolution does not quite converge on strategy pairs that would create the strongest stabilizing effect.

\section*{Discussion}

\subsection*{Main results}

Relative nonlinearity of competition (RNC), a classical dynamic coexistence mechanism, requires that species show different nonlinear responses to a shared resource, and that each species affects resource dynamics in a way that limits its own growth more than that of its competitors. We confirmed previous findings that this stabilizing effect readily arises in several population models with nonlinear density dependence, when an overcompensating species inhabits an environment together with a species that reacts significantly weaker to deviations of the total population size from the community's carrying capacity. 

However, the main finding of the present study is that this dynamically stable coexistence is evolutionarily stable only under fairly restrictive conditions. To arrive at this conclusion, we considered an evolutionary trade-off between growth at low resource availability and growth at high resource availability. 

In a first step, we used three classical population models (the Ricker, the Hassel, and the Maynard Smith and Slatkin model) to create families of density-dependence functions that follow this trade-off. Our analysis using adaptive dynamics theory (Fig.~\ref{figure: classical PIP}), as well as individual-based simulations (Figs.~\ref{figure: coexistence},\ref{figure: collapse},\ref{figure: localEvolution}), show that evolution generally tends towards a single evolutionary attractor, approximately located at the density-compensation strategy $b_\mathrm{cri}$ at which population dynamics switch from compensating (stable) to overcompensating (fluctuating) behavior. This is in line with previous findings that evolution tends towards the edge of stability \cite{Doebeli-EvolutionofSimple-1995, Rand-InvasionStabilityand-1995}, and interestingly, in this case even slightly beyond that edge. We speculated that the latter is a result of the introduction of density-independent mortality in our models, which may promote slight overcompensation to adjust for the additional mortality. Strongly overcompensating strategies that are required for the stabilizing feedback of RNC, however, were not favored in our initial analysis. 

In a second step, we systematically modified the evolutionary trade-off in such a way that the formerly stable intermediate density-compensation strategy was strongly reduced in fitness (eq.~\ref{eq: modified MSS}). These changes created disruptive selection and therefore evolutionary branching towards a pair of density-compensation strategies stabilized by RNC (Figs.~\ref{figure: pip-mod},\ref{figure: evolution-mod}). This demonstrates the possibility that RNC can evolve under an evolutionary trade-off between growth rates at low and high resource availability. However, achieving this outcome required considerable fine-tuning. We not only had to decrease the fitness of the previously favored compensating strategy, but at the same time had to ensure that evolutionary branching leads to a pair of density-compensation strategies that is dynamically stable. The density-compensation functions resulting from this fine-tuned trade-off look highly irregular compared to the unmodified MSS model (Fig.~\ref{figure: pip-mod}). Moreover, the location of the  formerly stable intermediate density-compensation strategy, and therefore, the necessary modification of the density-compensation functions, crucially depends on $r$ and $d$ (Fig.~\ref{figure: localEvolution}). We cannot imagine a feedback in nature that would lead to such a fine-tuned trade-off across a wide range of environments. Thus, based on our analysis, it seems rather unlikely that real trade-offs will meet the conditions required for the sympatric evolution or evolutionary stability of RNC. 

\subsection*{Relation to character displacement}

It is interesting to consider the relation of our results to character displacement, the process through which species reduce competition by diverging in their traits \cite{Losos-Ecologicalcharacterdisplacement-2000, Pfennig-CharacterDisplacementand-2010}. The reason why we do not observe character displacement in this study is that species cannot escape competition by changing their density-compensation strategy. Some specific combinations of $b$-values can coexist, which are those we identified in Fig.~\ref{figure: coexistence}b. Those pairs of density-compensation strategies partition the available fluctuating resources in a way that each species is stabilized at low relative frequency. In a generalized concept of the niche, one could say that each species has managed to find a disparate niche space by specializing either on high or on low levels of resource fluctuations. However, there is usually a "generalist" density-compensation strategy that can invade either of those "specialist" strategies, while it can not be invaded by them in return. This "generalist", which we identify as the evolutionarily stable strategy in Fig.~\ref{figure: localEvolution}, can exploit the niches of either of the two "specialist" species, and hence competitively excludes them (Fig.~\ref{figure: collapse}). 

\subsection*{Generality and scope of the results}

Our study could only explore a limited number of all possible families of functions that follow the ecologically motivated trade-off between growth rates at high and low resource availability, from which we departed in this study. Yet, given the fine-tuning necessary to achieve evolutionary stability, our results strongly suggest that only a very restricted subset of all these possible functional families allows evolutionarily stable coexistence by RNC. For RNC to be common, one would need a mechanism that explains why those functions in particular should be favored in nature. Further analysis of these questions could be conducted using models that explain density dependence from mechanistic assumptions, such as \cite{Johst-Fromindividualinteractions-2008}, but the drawback is that also these models require assumptions about trade-offs in species traits, although at a more basic level. Greater certainty could only be gained by analyzing trade-offs from empirical data, as reported in a study by Metcalf et al., which examines a trade-off in flowering time parameterized with real data \cite{Metcalf-Evolutionofflowering-2008}. Interestingly, Metcalf et al. do not find conditions that would allow the evolution of RNC in their plant system, in accordance with our results.

A limitation of our study is that our analysis is based on discrete-time models with non-overlapping generations and a fixed resource supply that does not show any lags. Previous research has suggested that RNC can also occur, and might even be particularly likely, in populations with non-overlapping generations and gradually regrowing resources \cite{Abrams-Whendoesperiodic-2004, Wilson-CoexistenceCyclingand-2005}, a situation we would expect for resources such as plant biomass, prey, or nutrients. At the moment, we cannot say anything about the evolutionary stability of RNC in these time-continuous systems. A detailed analysis of their evolutionary stability would be a valuable extension of this study. Another interesting extension could be to account for spatial structure in the evolving populations. It has been shown that density compensation coevolving with dispersal may lead to the evolutionarily stable coexistence of strategy pairs that differ both in density compensation and dispersal traits \cite{Johst-Evolutionofcomplex-1999}, but this may reflect a competition-colonization trade-off, rather than pure RNC. 

Our results by no means exclude the possibility of evolutionarily stable complex population dynamics and coexistence in fluctuating environments in general. What we have tested here is whether one specific coexistence mechanism, RNC, could evolve sympatrically, or be evolutionarily stable, in a spatially unstructured environment. We find that this does not seem particularly likely. An additional insight is that the evolution towards strong overcompensation with complex, oscillatory population dynamics through this mechanism seems rather unlikely as well. However, this does not exclude the possibility that complex dynamics could evolve through any other mechanism. The question of complex population dynamics, which have been observed both in nature and in experiments, has been a field of active research for many years \cite{Turchin-Complexpopulationdynamics-2003}, and many possible mechanisms have been proposed that are not excluded by this study. Therefore, if complex population dynamics are observed, it seems likely that one or several of these additional mechanisms are at work. 

Moreover, the finding that RNC is unlikely to evolve or be evolutionarily stable in sympatry does not mean that it could not emerge in other situations. Speciation, for example, may take place in isolated areas where species locally evolve different strategies (see, e.g., Fig.~\ref{figure: localEvolution}). When species from such a source pool emigrate to other areas, RNC may well play a role in temporarily stabilizing local species diversity. Similar opportunities may arise when rapid environmental change leaves species maladapted to the present environmental regime. It is an interesting avenue for further empirical and theoretical research to test whether the effects of these mechanisms are strong enough to substantially increase the expected or observed prevalence of coexistence by RNC.

\subsection*{Conclusions and outlook}

In conclusion, our results show that differences in density compensation may stabilize species coexistence on ecological time scales, but it seems generally rather unlikely that such coexistence can arise or be stable on evolutionary timescales. We believe that this distinction between dynamic and evolutionary stability, although noted before, is crucial for gaining a better theoretical understanding of the mechanisms behind diversity patterns in general, and of coexistence mechanisms in particular. An earlier study, for example, concluded that "the paradox of the plankton is essentially solved" after finding that a model of a planktonic community allows dynamically stable coexistence of more species than resources when population dynamics are chaotic \cite{Huisman-Biodiversityofplankton-1999}. Later, however, \cite{Shoresh-Evolutionexacerbatesparadox-2008} demonstrated that evolving resource partitioning may lead to a drastic breakdown of such dynamically stable diversity. Differences between evolutionary and dynamic stability have also been found or conjectured for a trade-off between maturation rate and birth rate \cite{Mougi-Evolutionofmaturation-2006}, for the aforementioned study of flowering decisions in plants \cite{Metcalf-Evolutionofflowering-2008}, and for a trade-off in a predator's handling time \cite{Kisdi-Evolutionofhandling-2006}. Following up on the last study, \cite{Geritz-Evolutionarybranchingand-2007} found that it was possible, but difficult, to construct trade-offs that allow evolution towards coexistence. Regarding the evolutionary stability of the temporal storage effect, \cite{Snyder-CoexistenceandCoevolution-2011, Abrams-Evolutionstorageeffect-2013, Mathias-Coexistenceandevolutionary-2013} report somewhat more favorable conditions, although some restricting conditions have to be met as well. Together, this shows that in many cases, there are significant differences between the coexistence mechanisms that would be beneficial for supporting maximum diversity in a community, and the coexistence mechanisms that we would expect to evolve. 
 
We believe that our study, as well as the other mentioned recent examples, show that there is still a surprising lack of knowledge regarding the interplay of dynamic and evolutionary mechanisms responsible for structuring ecological communities. A reason may be the lack of quantitatively reliable descriptions of trade-offs among a species' adaptive traits, including its life-history strategies, which makes comprehensive evolutionary analyses difficult. Nevertheless, we think that extending our theoretical understanding in this general direction is important, and may even be indispensable, for making quantitative predictions about the evolution of species diversity and biogeographic patterns. Only a combined analysis of community dynamics and evolutionary dynamics, as promoted also by the recent trend of eco-evolutionary approaches \cite{Dieckmann-Adaptivedynamicsand-2004, Johnson-emergingsynthesisbetween-2007, Matthews-Towardintegrationof-2011}, together with empirical data from both domains, may be able to provide a more conclusive answer to how different coexistence mechanisms contribute to ecological diversity across spatial and temporal scales.



%

\section*{Acknowledgements}
We would like to thank Thomas Banitz, Claudia Dislich, Simon Levin, and Gita Benadi for helpful comments. The models used in our analysis were implemented in Netlogo 4.1 and 5.0 \cite{Wilensky-NetLogo-1999}. Plots and statistical analysis were carried out with R version 3.0.2 \cite{RDevelopmentCoreTeam-RLanguageand-2014}. FH acknowledges support from the ERC Advanced Grant 233066. TM received funding from the European Community's Seventh Framework Programme (FP7/2007-2013) under grant agreement no 281422 (TEEMBIO). UD gratefully acknowledges financial support by the European Commission, the European Science Foundation, the Austrian Science Fund, the Austrian Ministry for Science and Research, and the Vienna Science and Technology Fund.



\end{document}